\documentclass[preprint, pra, amsmath, aps,showpacs]{revtex4}
\usepackage[dvips]{color}
\usepackage[usenames]{xcolor}
\newcommand{\bea}{\begin{eqnarray} }
\newcommand{\eea}{\end{eqnarray}}
\newcommand{\bean}{\begin{eqnarray*}}
\newcommand{\eean}{\end{eqnarray*}}
\newcommand{\nn}{\nonumber \\}
\def\B{{\bf B}}

\def\btimes{~{\bf \times}~}
\def\bnabla{{\bf \nabla}}
\def\bcdot{~{\bf \cdot}~}
\newcommand{\lbm}{\left\lbrack}
\newcommand{\rbm}{\right\rbrack}

\def\av#1{\left\langle #1 \right\rangle}
\def\od#1,#2{\frac{d#1}{d#2}}
\def\odz#1,#2{\frac{d^2#1}{d{#2}^2}}
\def\pd#1,#2{\frac{\partial #1}{\partial #2}}
\def\pdz#1,#2{\frac{\partial^2 #1}{\partial {#2}^2}}

\def\pdd#1,#2{\frac{\partial^3 #1}{\partial {#2}^3}}
\def\pdv#1,#2{\frac{\partial^4 #1}{\partial {#2}^4}}
\def\pdzz#1,#2,#3{\frac{\partial^2 #1}{\partial {#2}\partial{#3}}}

\def\eq#1{Eq.~(\ref{#1})}

\usepackage{graphicx}% Include figure files
\usepackage{dcolumn}% Align table columns on decimal point
\usepackage{bm}% bold math

\begin{document}

%\preprint{\small {\color{red} Change-marked copy (in red)}}
%\draft
\bibliographystyle{unsrt}
%\preprint{\small }
%
% -------------------------------------------------------------
%Prospects of negative triangularity tokamak \\for advanced steady-state confinement
%Prospects of negative triangularity configuration  \\for advanced tokamak scenario in steady-state confinement
%\title{Prospects of advanced steady-state confinement of fusion plasmas \\by negative triangularity tokamak}
\title{Prospects of negative triangularity tokamak  \\for advanced steady-state confinement of fusion plasmas}
%
% -------------------------------------------------------------
%
\author{Linjin  Zheng,\email{lzheng@austin.utexas.edu} M. T. Kotschenreuther, F. L. Waelbroeck, 
M. E. Austin, W. L. Rowan, P. Valanju, and X. Liu }
\affiliation{Institute for Fusion Studies,
University of Texas at Austin,
Austin, TX 78712}
\date{\today}

\begin{abstract}

The steady-state confinement,  beta limit, and divertor heat load are among
the most concerned issues for toroidal confinement of fusion plasmas. In this work, we show that
the negative triangularity tokamak has promising prospects to address these issues. 
We first demonstrate that the negative triangularity tokamak generates the
filed line rotation transform more effectively. This brings bright prospects for the
 advanced steady-state tokamak scenario.
Given this, the stability and confinement features of negative triangularity tokamak 
are investigated. We point out  that the negative triangularity configuration with
a broad pressure profile
is indeed more unstable for low-$n$ magnetohydrodynamic modes than the positive triangularity case
so that the H-mode confinement can hardly be achieved in this configuration, 
where $n$ is the toroidal mode number. 
Nevertheless, we found that the negative triangularity configuration
with high bootstrap current fraction, high poloidal beta,
and peaked pressure profiles  can achieve higher normalized beta  
for low-$n$ modes  than the positive triangularity case. 
In a certain parameter domain, the normalized beta   can reach about  twice 
the extended Troyon limit,
while the same computation indicates that the positive triangularity 
configuration is indeed constrained by the Troyon limit. 
  This shows that the negative triangularity tokamaks are not only favorable for divertor design to avoid the edge localized modes but also can 
have promising prospects for advanced steady-state confinement of fusion plasmas in high beta.

\end{abstract}

\pacs{52.53.Py, 52.55.Fa, 52.55.Hc}

\maketitle
%\bibliography{zhengbib.bib}

 \section{introduction}

The steady-state confinement,  beta limit, and divertor heat load  are among
the most concerned issues for toroidal confinement of fusion plasmas. These issues
depend sensitively on the equilibrium configurations, positive or negative triangularity.
In conventional tokamaks, including ITER \cite{iter}, the positive triangularity configuration
is chosen. This is mainly because the H-mode confinement --- an operating mode with high energy confinement --- 
can be achieved in this type of configurations \cite{hmode}. This seems to be favorable
for the stability in high beta confinement. Here,
beta ($\beta$) represents the ratio between plasma and magnetic pressures.
However, it is gradually realized that
the divertor heat load is a major concern for the  H-mode confinement. 
This is because the H-mode confinement is often tied to
 the damaging edge localized modes (ELMs) \cite{hmode}. ELMs can discharge particles and heat into the scrape-off layer and subsequently to the divertors. The divertor plates can be  damaged by such a discharge.

The negative triangularity tokamak is  proposed as a possible solution  \cite{tcv,kikuchi14,d3d}. 
When the divertor configuration is moved from the high field side in the
positive triangularity case to the low field side in the negative triangularity case,
a larger separatrix wetted area is achieved due to the larger major radius of
divertor location. With the divertor in the low field side, 
a larger room is yielded for divertor engineering design. 
However, as discussed  in Refs. \onlinecite{yu15,ziaea,zint,zrot}, the stability beta limit is a concern for negative triangularity tokamaks.  One can hardly achieve the
H-mode confinement in the negative triangularity configuration. 

However, there is an important development.  Recent TCV and DIII-D experiments 
found that the low (L) mode discharges in the negative triangularity experiments can reach about the same level 
of normalized beta  as the H mode confinement in  the positive triangularity tokamaks  \cite{tcv,d3d}.  Here,  the normalized beta    $\beta_N^{geo}$  is defined as  $\av{\beta}/ I_N$,  where $\av{\beta}$ is 
the ratio of the volume-averaged plasma pressure to the toroidal vacuum magnetic pressure at the 
geometric center of plasma column, $I_N = I/aB$ (MA/m/T) is the normalized toroidal current,
$I$ is the toroidal plasma current, $B$ is the vacuum toroidal magnetic field at the 
geometric center,  $a$ is the minor  radius. This definition is used in DIII-D experiments \cite{betand3d}.
It is slightly different from the original definition $\beta_N^{Troyon}$ in 
Refs.  \onlinecite{troyon1} and \onlinecite{troyon2}, in which $\av{\beta}^{Troyon}$ is defined as 
the ratio of the volume-averaged plasma pressure to the volume-averaged  toroidal magnetic pressure.
Since they are L-mode discharges, there are no ELMs.  Furthermore, the turbulence level is also  found to be 
considerably low  in these discharges  \cite{tcv,d3d,zint}.
This has further stimulated the interest in  the negative triangularity tokamaks
and motivates us to further explore the concept of negative triangularity tokamaks. 

In the equilibrium, we focus on examining the safety factor value difference between
the positive and negative triangularity tokamaks.  
Using the analytic Solov\'ev equilibrium \cite{sol}, DIII-D-homologous configuration, 
and the advanced steady-state confinement scenario with high bootstrap current fraction, high poloidal beta,
and peaked pressure profiles or the case with internal transport barrier,
we found that the negative triangularity tokamak is more effective in generating the
filed line rotation transform. 

Note that both analytic Solov\'ev equilibrium and DIII-D homologous configurations 
yield lower safety factors in negative triangular configurations. This leads us 
to promote an advanced steady-state scenario with high bootstrap current fractions.
We found that even when the bootstrap current fraction reaches as high as $95\%$, the
achievable case with the numerical equilibrium codes, the safety factor value in the edge region remains
reasonably low. This opens  promising prospects for the steady-state confinement
in the negative triangularity configuration. 

In the pressure profile optimization, we note that 
 the H mode can hardly be achieved in the negative triangularity configuration.
This is because the negative triangularity configuration with a broad pressure profile
is usually more unstable for low-$n$ modes than the positive triangularity case.
 We also note that  the numerical simulation in Ref.  \cite{nash}  shows that 
 the negative triangularity configuration cannot well
confine trapped ions. The experimental observation also indicates that
the negative triangularity configuration needs higher density for
divetor detachment \cite{detach}. 
Furthermore, as we will see  that the high local $q$ region is narrower in the
negative triangularity configuration, which is shown in Ref. \onlinecite{zxst}
to be unfavorable to the X-point stabilization of edge modes.  
All of these point to that the negative triangularity configuration
favors the peaked pressure profiles or the case with internal transport barrier.
A further reason that leads us to consider the pressure profile to be more 
peaky than that of the L-mode type is due to the specific stability properties discussed next.
 
Based on these interesting features, the magnetohydrodynamic (MHD) stability of  the
advanced steady-state confinement scenario with high bootstrap current fraction, high poloidal beta,
and peaked pressure profiles or internal transport barrier in the negative
triangularity configuration is investigated.
We found that
the advanced negative triangularity scenario can achieve a higher normalized beta  
for low-$n$ modes than the positive triangularity case \cite{betand3d}. 
In a certain parameter domain, the normalized beta  limit can reach about $8$ $l_i (I/aB)$ for low-$n$ modes, about twice 
the extended Troyon limit for conventional positive triangularity tokamaks. 
This good stability property is attributed to the equilibrium  feature
 that the steep pressure gradient appears in the core region where the magnetic shear is negative,  while  the
 safety factor  near the edge region remains to be low.  
This leads us to conclude that the negative triangularity tokamaks 
have  promising prospects  for advanced steady-state confinement of fusion plasmas in high beta.

The paper is arranged as follows: In Sec.~\ref{sec.eq}  the negative triangularity equilibria are explored 
to show its effectiveness in generating the field line rotation transform.  The Solov\'ev equilibrium,
the DIII-D-homologous equilibria, and the advanced steady-state scenario are examined. 
In Sec.~\ref{sec.st}, the MHD stability both of the DIII-D-homologous equilibria and
the advanced steady-state scenario is studied.
The conclusions and discussion are given in the last section.

 %%%%%%%%
 \section{Equilibrium: low safety factor feature of negative triangularity configuration}
%%%%%%%

\label{sec.eq}

One of the key challenges for tokamaks is  steady-state  confinement.
The toroidal confinement requires the field-line rotation transform. In tokamaks,
it is realized by inducing the toroidal plasma current using the transformer 
principle with plasma working as the secondary coil. Due to the current saturation 
of the primary coil, one cannot hold the tokamak Ohmic current steadily. For steady-state 
confinement, the bootstrap current or other current drive means are resorted to.
We found  that the negative triangularity configuration is
more effective in generating the field-line rotation transform, i.e., 
 reducing the safety factor.

We consider the axisymmetric tokamak configuration. In this case, the magnetic field can be expressed
as follows
\bea
\B&=& \bnabla\phi\btimes\bnabla\chi +f(\chi)\bnabla\phi,
\label{bb00}
\eea
where $f$ denotes the poloidal current density flux. One can arbitrarily introduce a poloidal angle
$\theta_{eq}$ here.
Introducing the so-called flux coordinates, for example, the PEST coordinates \cite{pest},
\eq{bb00} is transformed to
\bea
\B&=& \bnabla\phi\btimes\bnabla\chi +q(\chi)\bnabla\chi\btimes\bnabla \theta_p.,
\label{bb}
\eea
where $\phi$ is the axisymmetric toroidal angle,  $\chi$ is the poloidal magnetic flux, 
  $q(\chi)$ denotes the safety factor,
 and $\theta_p$ is the poloidal angle. In the  so-called PEST coordinates, one has
\bea
\theta_p&=& \frac f{q}\int_0^{\theta_{eq}} d\theta_{eq} \frac {{\cal J} }{X^2},
\nn
q&=& \frac f{2\pi}\oint d\theta_{eq} \frac {{\cal J} }{X^2},
\label{qq}
\\
{\cal J} &=& \frac 1{ \bnabla\phi\btimes\bnabla\chi\bcdot \bnabla\theta_{eq}}.
\label{jac}
\eea
One can derive the Grad-Shafranov equation for an axisymmetric system.
In cylindrical coordinates ($X,Z,\phi$), it  can be expressed as
\bean
X\pd, X\frac1X\pd \chi,X+\pdz \chi,Z& =& -\mu_0P_\chi' X^2- ff_\chi', 
\eean
where  $X$ is the major radius, $Z$ is the height,  $\mu_0$ is the magnetic constant,
 $P$ is the pressure,
  and the prime denotes the derivative with respect to $\chi$. 

We examine both  the Solov\'ev analytical equilibrium  and numerical equilibria.
For numerical equilibria,  we use the VMEC code
 as the Grad-Shafranov equation solver \cite{hirshman}, 
 with the bootstrap current  being included from the Sauter formula \cite{sauter}, to extensively 
construct the tokamak equilibria. 
The toroidal current profile in the Grad-Shafranov equation  is determined self-consistently by specifying the density 
and temperature profiles. When the temperature and density profiles are given, the pressure is determined. 
Also, since the plasma resistivity depends on  temperature, the Ohmic current is then determined by the given temperature profile.
The bootstrap current is also determined by the given density and temperature profiles. 
By iteration, the amount of Ohmic current  is varied to yield a specified total current (bootstrap plus Ohmic). 
In this process, the amount of the Ohmic current is varied consistently with the neoclassical conductivity.
Both DIII-D-homologous and advanced steady-state scenarios are investigated numerically. 
  The DIII-D-homologous
equilibria have been used for stability analyses as reported in Refs.  \onlinecite{ziaea} and \onlinecite{zint}. 
Here, we focus on discussing
their safety factor features.

 %%%%%%%%
 \subsection{Solov\'ev equilibrium}
%%%%%%%

In this subsection, 
we  examine the Solov\'ev equilibrium  \cite{sol}. 
 The Solov\'ev equilibrium solution also allows us to explore
the triangularity effects on the safety factor semi-analytically.
  In the Solov\'ev equilibrium, one assumes that
\bea
 -\mu_0P_\chi' =a~~~ \hbox{and}~~~ -ff_\chi'=bX^2.
 \label{pfab}
 \eea
We consider the exact Solov\'ev equilibrium solution as  given as follows 
\bea
\chi &=&\lbm (b+c_0)X_0^2+c_0(X^2-X_0^2)\rbm \frac{Z^2}2 +\frac18(a-c_0)(X^2-X_0^2)^2,
\label{chi0}
\eea
where $X_0$ denotes the major radius of magnetic axis, $a$, $b$, and $c_0$ are  constant parameters.
In order to determine the cross section and for the safety factor calculation, we first determine
the separatrixes of the solution in \eq{chi0}. They are obtained by the
stationary points of $\chi$:
\bean
\pd \chi, Z&=&0 ~~\to~~~ X^2-X_0^2 = -\frac{b+c_0}{c_0}X_0^2,
\nn
\pd \chi, {(X^2-X_0^2)}&=&0 ~~\to~~~ Z^2=-\frac12\frac{a-c_0}{c_0} (X^2-X_0^2). 
\eean
From them, one can determine the parameters
\bea
\frac a{c_0}=1-2\frac{Z^2_s}{ X_s^2-X_0^2},
~~~
\frac b{c_0}&=& -1 - \frac{X_s^2-X_0^2}{X_0^2},
\label{xpn}
\eea
where $(X_s,Z_s)$ are the X-point coordinates.
Using \eq{xpn}, let us now consider the case with a quasi-mirror symmetry of positive and negative triangularity cases,
i.e., $X_0^2 -X_{sp}^2 = X_{sn}^2-X_0^2$, where the subscripts $p$ and $n$ 
are introduced to denote the positive and negative triangularity cases, respectively.
One therefore has 
\bean
\frac {a_n}{c_{0n}} =2 -\frac {a_p}{c_{0p}},
~~~  \frac {b_n}{c_{0n}} =-2 -\frac {b_p}{c_{0p}}.
%\label{pn2}
\eean

  We have extensively examined the general  feature of safety factor in the Solov\'ev equilibria. Here, let us  
discuss the case with the DIII-D-like geometric parameters. 
We choose $X_0=1.67$, $X_{sp}=1.05$,  $Z_{sp}=1$, and the beta at the magnetic axis  $\beta_0=0.03$. 
In this case, the quasi-mirror-symmetric positive and negative triangularity Solov\'ev equilibria can be determined. 
The  equilibrium cross sections are plotted  in Fig. \ref{f2surf}, with the $\beta$ and $f$ profiles
given respectively in Figs. \ref{f2pr} and \ref{f2gt}.  From \eq{chi0} one can see that $c_0^2$ can be absorbed into
the definitions of $a$ and $b$. We therefore choose $c_0=1$ for simplicity. 
$a= -0.1860$ and $b=-1.6047$ in the negative triangularity case
and $a= 2.1860$ and $b=-0.3953$ in the positive triangularity case.
Here, it should be pointed out that in determining $f$ from $b$ there is an integration constant,
which is actually related to the magnitude of toroidal field, The integration constant is therefore
used to scale the beta value at the magnetic axis. 

The safety factor can be computed for the equilibria shown in Fig. \ref{f2surf} respectively 
for positive and negative triangularity cases. The results are plotted in Fig. \ref{f2qsf}. 
Because at the X point, $|\bnabla\chi|$ vanishes. The Jacobian in \eq{jac} becomes
infinite. Consequently, as is well known the surface-averaged safety factor becomes infinite
on the last closed flux surface as shown in Fig. \ref{f2qsf}. From the definition of the safety factor in \eq{qq}
one can see that the safety factor is a surface-averaged quantity. 
Noting that the Jacobian only becomes singular at the X points,
one can expect that the integrand in the definition of safety factor, \eq{qq}, is not singular
everywhere. This leads us to plot out the local safety factor profiles, $q_{local} = {\cal J} f/X^2$,
in Fig. \ref{f2qloc}. 

Let us now discuss some interesting features of the safety factor 
in the positive and negative triangularity cases. First, from Figs. \ref{f2qsf} and \ref{f2qloc} one can see
that both of the  surface-averaged and local safety factors in the negative triangularity case are
lower than that in the positive triangularity case. This is also seen
in our VMEC equilibrium calculations with the bootstrap current taken into account to be described later on.

Next, as pointed out  in our recent work in Ref. \onlinecite{zxeq}, the equilibrium with
X point on the plasma edge occurs only if the included angle of plasma segment is 
 90 degrees.  Otherwise, the X point is present in the vacuum region and the plasma-vacuum
 interface becomes a hyperbola near the X point.
  A thin plasma edge layer needs to be truncated off to form a hyperbola type
of boundary with the X point located in the vacuum region. From Figs. \ref{f2qsf} and \ref{f2qloc}
one can see that such a truncation affects the edge $q$ more in the negative triangularity
case than in the positive one. Furthermore, 
from Fig. \ref{f2qloc} one can see that the surface-averaged $q$ as shown in Fig. \ref{f2qsf}
 may not be a relevant quantity 
to describe the X-point effects on the MHD modes. The local $q$ depends on the poloidal location.
This leads Ref. \onlinecite{zxst} to introduce
the dual-poloidal-region $q$ description.
In this description,  the X-point stabilization is more extensive 
 in the positive triangularity case than in the negative triangularity case 
 since its high $q$ region is more extensive in the radial direction. 
In Ref. \cite{nash} it is shown that the trapped ions can hardly be confined in the
negative triangularity configuration. Therefore, the negative triangularity 
configuration has a stronger radial electric field. Nonetheless,
the H-mode can hardly be achieved in the negative triangularity case.
Also, the density for the divertor detachment  in the negative triangularity case is required to be higher \cite{detach}.
It indicates that the edge transport in the L mode confinement in the negative triangularity
configuration is still strong.
With the X-point stabilization effects discussed in Ref. \onlinecite{zxst},
one may explain why the H mode can occur in the positive triangularity
case, instead of the negative one.
This is because the positive triangularity configuration contains
a larger high local $q$ region. This leads the MHD modes to be stabilized in
the pedestal region. This leads us to conclude that the
equilibrium with the internal transport barrier or peaked pressure profile
is preferred in the negative triangularity case. This is also consistent 
with the experimental observations in TCV and DIII-D \cite{tcv,d3d} that
the negative triangularity configuration favors the L mode confinement.
We push further to consider the case with an even more peaked pressure profile. 

In passing, it is also interesting to point out that there is a limitation to generating the
Solov\'ev equilibria with  negative triangularity. This is because
both $P_\chi'$ in \eq{pfab}  and $\chi$ in \eq{chi0} by definitions depend on the parameter $a$. To ensure
the pressure profile is monotonically decreasing in the radial direction, 
the elongation cannot be too small. For example,
the solution does not exist if $Z_s =0.9$ in the Fig. \ref{f2surf} case.
The negative triangularity case in Fig. \ref{f2surf} is achieved by a negative $a$ and monotonically decreasing
$\chi$, for the parameters mentioned above. This does not mean that
the small elongation equilibrium does not exist in the negative triangularity case
since a homogeneous solution can be added in. How this phenomenon affects
the stability of negative triangularity configuration without a large elongation 
remains to be investigated. Coincidentally, the advanced steady-state scenario 
to be introduced later on is indeed further elongated.

 %%%%%%%%
 \subsection{DIII-D-homologous equilibria}
%%%%%%%
\label{subsec.eqd3d}

In this subsection,  we  discuss the DIII-D-homologous negative triangularity equilibria. The equilibria
have been used for stability analyses  in Refs. \onlinecite{ziaea} and \onlinecite{zint}. 
Here, we focus on discussing the safety factor feature of the negative triangularity configuration
as compared to the positive one.

 The experimental results for DIII-D  negative triangularity discharges were reported in Ref. \cite{d3d}. 
 As explained in Refs.  \onlinecite{ziaea} and \onlinecite{zint},
the  experimental equilibrium is numerically reconstructed by the EFIT code \cite{efit}.
 The toroidal 
magnetic field at the geometry center  is $B_T=2$T, the total current is fixed to be $I=0.9$MA, and the major and minor radii are respectively $R=1.67$m and  $a=0.6$m.
To extend the  parameter domain outside the experimental values in DIII-D,  we numerically  construct  
the DIII-D-homologous equilibria with  positive, zero, and negative triangularity:
$\delta = 0.4$, $0.0$, and $-0.4$. The L-mode pressure profile is used to simulate DIII-D experiments.
The procedure has been reported in   Refs.  \onlinecite{ziaea} and \onlinecite{zint} and
also outlined at the beginning of this section.
The typical cross sections of the positive and negative triangularity configurations of DIII-D-homologous equilibria
are given in Fig. 1 of Ref.    \onlinecite{zint}.  
The main equilibrium parameters are summarized in Table \ref{table0}, in which
$\delta$ denotes triangularity, $\beta_N^{geo}$ is the normalized  beta as used in the DIII-D experiments,
$\beta_N^{Troyon}$ is the normalized beta introduced by Troyon,  $\av{\beta}$ is the volume-averaged
beta,  $l_i$ is the inductance per unit length, and $f_{bs}$ represents the bootstrap current fraction.
 \begin{table}[h!]
  \begin{center}
    \caption{Equilibrium parameters}
    \label{table0}
    \begin{tabular}{|c||c|c||c|c|}
    \hline
   Case  & PT1 & PT2 & NT1 & NT2 
     \\
     \hline
     \hline
$\delta$& 0.4&0.4&-0.4&-0.4
\\
\hline    
$\beta_N^{geo}$& 3.79&4.74&2.68&3.38
\\
\hline    
$\beta_N^{Troyon}$& 3.35&4.23&2.73&3.47
\\
\hline    
$\av{\beta}(\%)$& 2.84&3.55&2.01&2.53
\\
\hline    
$l_i$& 0.705&0.600&0.806&0.702
\\
\hline    
$f_{bs}$(\%)& 53&66&43&53
\\
\hline    
    \end{tabular}
  \end{center}
\end{table}

To show the safety factor features of positive and negative triangularity configurations,
the pressure (or $\beta$ normalized by the  magnetic field at the axis) and safety factor profiles 
for positive and negative triangularity cases given in Table \ref{table0} are plotted
in Figs. \ref{dptpq} and \ref{dntpq},  respectively. 
Again,  by comparing Figs. \ref{dptpq} and \ref{dntpq} one can see that
the negative triangularity configuration leads to a lower safety factor value 
$q$, especially near the edge.  
 Note that both positive and negative triangularity equilibria are created with
the same total toroidal current and toroidal magnetic field. The low $q$ feature leads us to
 explore the possibility of steady-state confinement in the negative triangularity
 configuration in the next subsection.

 %%%%%%%%
 \subsection{Advanced steady-state confinement scenario}
%%%%%%%
\label{subsec.eqadv}

 In this subsection, 
we  examine the advanced steady-state scenario in the negative triangularity
configuration \cite{ziaea}. As shown in the Solov\'ev equilibrium and the DIII-D-homologous cases, 
the negative triangularity configuration is shown to be more effective in creating the field line rotation transform  
 as compared to the positive triangularity configuration
 with the same given toroidal current and toroidal magnetic field. This motivates
 us to explore the  advanced  steady-state scenario  with a high bootstrap current fraction in the negative triangularity
 configuration.  
  
 Since this
 is  new, we are no longer bound to the DIII-D  configurations.  Instead, we consider the configuration with the elongation $\kappa=2$ and aspect ratio being  $3$ as often employed for steady-state tokamak studies. 
 We mainly investigate the negative triangularity 
 configurations, while a comparison with the positive case is also made. The typical cross-sections for positive ($\delta=0.4$) 
 and negative ($\delta=-0.4$) triangularity configurations are shown in Fig. \ref{advcross}.
 
 In further exploring the equilibrium parameter
domain, we found that  the negative triangularity favors the peaked pressure profiles; 
while the positive triangularity the broad pressure profiles. This leads us to consider the 
 cases with high bootstrap current fraction, high poloidal beta,
and peaked pressure profiles. 
We consider five negative triangularity cases with $\delta = -0.4$.  Their equilibrium parameters are given in Table
\ref{table1} and the pressure and safety factor profiles are given in Figs. \ref{advp} and \ref{advq}, respectively. The equilibrium sequences 
are generated by raising up the center pressure. In these cases, the magnetic field is $2$T, the total toroidal current
is $1.2$MA, and the minor radius is $0.6$m. From Table \ref{table1} one can see that the bootstrap current fraction is very high in these
equilibrium cases. We cannot generate equilibrium cases with even higher bootstrap current fraction because the loop between 
the Grad-Shafranov equation solver and the subroutine for computing bootstrap current tends to become hard to converge. 
The Mercier stability status is also given in Table \ref{table1}, which will be discussed in subsection \ref{subsec.stadv}.
They are the cases with an internal transport barrier.
  \begin{table}[h!]
  \begin{center}
    \caption{Equilibrium parameters}
    \label{table1}
    \begin{tabular}{|c||c|c|c|c|c|}
    \hline
     & \textbf{Case 1} & \textbf{Case 2} & \textbf{Case 3} & \textbf{Case 4} & \textbf{Case 5}
     \\
     \hline
     \hline
$\beta_0$(\%)& 11.5&11.9&12.3&12.7&13.1 
\\
\hline    
$\av{\beta}(\%)$& 3.60&3.86&4.12&4.41&4.60 
\\
\hline    
$\beta_N^{geo}$& 3.60&3.86&4.12&4.41&4.60 
\\
\hline    
$\beta_N^{Troyon}$& 3.71&3.98&4.27&4.58&4.79 
\\
\hline    
$l_i$& 0.708&0.674&0.639&0.602&0.585 
\\
\hline    
$4(I/aB)$&4&4&4&4&4 
\\
\hline    
$4l_i(I/aB)$& 2.84&2.68&2.56&2.40&2.34 
\\
\hline    
$f_{bs}$(\%)& 80&82&85&88&95 
\\
\hline  
\hline    
Mercier& (0.20,0.35)&(0.24,0.33)&stable&stable&stable 
\\
\hline    
    \end{tabular}
  \end{center}
\end{table}

Again,  it is important to point out that the negative triangularity configuration
is so effective in generating the field line rotation transform that
 the low edge $q$ feature is insensitive to raising 
the bootstrap current fraction.  This can be seen in Fig. \ref{advq}.
In particular, in this figure
one can see that the $q$ value near the plasma edge remains to be
low even for the case with the bootstrap current fraction being $95\%$.

%%%%%%%%%%%%%%%%%%%%%%%%%%
  \section{The advanced stability features of negative triangularity tokamaks}
 \label{sec.st}

  In this section, we investigate the MHD stability of the negative triangularity tokamaks.
 Both the DIII-D-homologous equilibria and the advanced steady-state confinement scenario
  are investigated.  
 In Sec.  \ref{sec.eq}, we have shown that the negative triangularity configuration
is  more effective in generating the field line rotation transform. 
We will show that the negative triangularity configuration
can also be good for stability.
The results are reported in several conferences, including the 27th IAEA Fusion Energy Conference  \cite{ziaea}.
 The low $n$ mode stability features of DIII-D-homologous equilibria were reported in Refs. \onlinecite{zint} and
 \onlinecite{zrot} to show the comparison with 
 the intermediate $n$ modes and to study the rotation and diamagnetic effects. 
 To compare with the advanced steady-state confinement scenario, the DIII-D-homologous
 cases are reviewed here with
 further details reported. 
 
For stability, we  investigate the low-$n$ MHD stability using the AEGIS code \cite{Zheng2006748}. AEGIS is a linear ideal MHD stability code. 
It uses the Fourier decomposition  in the poloidal direction and independent solution decomposition  in the radial direction. 
Adaptive shooting  is used to obtain the independent solutions. 
AEGIS has been developed for several years. Here, let us briefly discuss its benchmarks with
other MHD codes.
The benchmarks with other major MHD codes in this field were performed when it was 
developed \cite{Zheng2006748}, for example, GATO \cite{gato}. It has been benchmarked with other codes in the last decade. Very recently, its comparison with MISHKA for 
Alfv\'en calculation is also shown to give a good agreement \cite{oliver}. Nevertheless, since rather high beta stability is found in the
AEGIS numerical computation for negative triangularity tokamaks, we employ  the DCON code \cite{glasser-a97} as a double
check. The results obtained in the AEGIS calculations are found to be consistent with the DCON results.
Besides, we also study the localized mode stability: the Mercier criterion and ballooning mode stability
using the codes in the DCON package.
This is the usual procedure for developing  a new concept.  The conformal wall is used in our calculations.

 %%%%%%
  \subsection{Stability of DIII-D-homologous equilibria}
 %%%%%%%
 \label{subsec.std3d}

To compare with the advanced steady-state confinement scenario  in Subsection \ref{subsec.stadv}, in this subsection, 
we outline the MHD stability results of DIII-D-homologous equilibria. The equilibria have been
described in subsection \ref{subsec.eqd3d}. They are based on the DIII-D experimental g-file
with the extrapolation to the positive and negative configurations with various beta values. 
Considerable results were reported in Ref.  \onlinecite{zint}. Here, we 
add more details and some further results.  

We first study the DIII-D negative triangularity discharges using the q-file. 
 Our stability analyses using the AEGIS code, together with DCON,  confirm that the equilibrium with the normalized beta  of 2.6 achieved in DIII-D experiments is
stable against the  $n=1$ MHD kink modes with critical wall position $b = 1.18$ and $n=2,3$ modes with the critical wall position
$1.05$.  DIII-D is not originally configured for negative triangularity discharges, the last closed flux surface touches the inner wall
and the X-points are  close to the outer part of the wall in the negative triangularity experiments.
Given that a large volume of magnetic flux touches the  wall and the parallel mobility of electrons, as an estimate we believe 
the case is  not far from the fixed boundary scenario. This is different from the  DIII-D positive triangularity discharges, where the wall is much further from the
plasma. 
 Note that  the inner wall
becomes effectively a limiter and the outer part of the wall distance is about 1.2 with the X-points close to the wall. Our results about the critical wall positions
 seem to be  consistent with the experimental observation.

To investigate the DIII-D-homologous numerical equilibria as described in Subsection \ref{subsec.eqd3d}, we also use the AEGIS code, together with DCON.
It is found that for the same given  density and temperature profiles, 
the positive triangularity case is  best,  the zero  triangularity case remains in the middle, and the negative  triangularity case is worst for $n=1$ MHD mode
stability. While the normalized beta  limit is reduced considerably  in the negative  triangularity cases,  it nonetheless still remains reasonably high.
This can be seen from the numerical results summarized as follows \cite{zint}. 
For each normalized beta  $\beta_N^{geo}$  we compute the critical wall positions for different triangularity cases. 
 The system is more stable if the critical wall position is larger. 
 Table \ref{table00}
 shows the critical wall position $b_c$ versus the normalized beta  $\beta_N^{geo}$ for three different 
triangularity cases,
  Actually, the profiles given in Figs. \ref{dptpq} and \ref{dntpq} 
are related, respectively, to the stability conditions in Table \ref{table00}
: The profiles in Fig. \ref{dptpq} correspond
to the positive triangularity cases with ($b_c=1.51$, $\beta_N^{geo}= 3.79$) and ($b_c=1.21$, $\beta_N^{geo}= 4.74$);
the profiles in Fig. \ref{dntpq} correspond to the negative  triangularity cases with ($b_c=1.86$, $\beta_N^{geo}= 2.68$)
and ($b_c=1.05$, $\beta_N^{geo}= 3.38$).
 \begin{table}[h!] 
  \begin{center}
    \caption{Critical wall position versus $\beta_N^{geo}$}
    \label{table00}
    \begin{tabular}{|c||c|c|c||c|c||c|c|c|}
    \hline
     Case & PT1 & PT2 & PT3& OT1  &OT2  &NT1 & NT2 & NT3 
     \\
     \hline
     \hline
$\delta$&0.4&0.4&0.4& 0.0&0.0&-0.4&-0.4&-0.4
\\
\hline    
$\beta_N^{geo}$&3.79&4.36   &4.74  &3.11&3.32   &2.68 &3.14& 3.38 
\\
\hline  
$b_c$&1.51&1.28&1.21& 1.46&1.31& 1.86&1.14&  1.05
\\
\hline    
    \end{tabular}
  \end{center}
\end{table}

.

In Ref. \onlinecite{zint}, we also studied the intermediate $n$ mode stability in  the negative triangularity case
with the DIII-D-homologous equilibria. 
It is found that although the low $n$ modes are more unstable in the negative triangularity case,
 the intermediate-n modes with a resistive wall are  more stable in the negative triangularity case than in the positive triangularity case \cite{zint}. This is consistent with the experimental observations of the lower level of turbulent transport in the negative triangularity case in TCV and DIII-D experiments \cite{tcv,d3d}.

In Ref. \onlinecite{zrot}, we also studied the rotation and diamagnetic drift effects 
on the $n=1$ modes in  the negative triangularity case
with the DIII-D-homologous equilibria. 
It is found that although the low $n$ modes are more unstable in the negative triangularity case,
the rotation and diamagnetic drift stabilization effects  are more effective in the negative triangularity case 
than in the positive triangularity case \cite{zrot}.

 %%%%%%
  \subsection{Stability property of the advanced steady-state confinement scenario}
 %%%%%%%
  \label{subsec.stadv}

In this subsection, we study the  stability of the negative triangularity equilibria 
in the advanced steady-state scenario described in subsection \ref{subsec.eqadv}.
We have shown that the negative triangularity equilibria 
in the advanced steady-state scenario  with high bootstrap current fraction, high poloidal beta,
and peaked pressure profiles have a favorable  equilibrium property 
in that it can effectively generate the field line rotation transform,
In this subsection, we show that the advanced steady-state scenario 
also has outstanding stability properties.
The tendency  in the DIII-D-homologous equilibria with a low bootstrap current fraction that the positive triangularity
 configuration is more stable, as described  in Table \ref{table00} in subsection \ref{subsec.std3d}, is reversed. 
 The negative triangularity configuration in this scenario becomes more stable than the positive triangularity case. 
 
We first study the cases with an internal transport barrier as shown in Fig. \ref{advp}.
One can first see the favorable stability features of the advanced steady-state scenario from the Mercier criterion
shown in Table \ref{table1}. In Table \ref{table1},
 the unstable regions in various cases are indicated by the normalized poloidal flux range.
The Mercier interchange modes become largely stable as beta increases,   possibly because the Grad-Shafranov shift increases
or the negative magnetic shear region becomes wider.   Here, it is noted that the Mercier stability in Cases  3, 4, and 5 excludes the vicinity of the magnetic axis. The Mercier instabilities usually appear at the vicinity of the magnetic axis. In  three cases, they are roughly located inside $\chi<0.002$. They may only
cause the local pressure profile to flatten at the magnetic axis.

All five equilibrium cases in Table \ref{table1} are no-wall stable 
to the $n=1,2,3$ external kink modes. We have specially studied  Cases 4 and 5 since their  beta values are high. The $n=4$ modes are also no-wall stable for Case 4. The critical wall position for $n=5$ modes is $2.53$, which is big 
noting the geometry parameters shown in Fig. \ref{advcross} (the wall is about touching the axisymmetric axis). Since the numerical equilibrium for Case 5 is marginally acceptable in convergency, 
we cannot pursue higher $n$ mode calculations. 
 Since the self-consistent equilibrium computation involves  both the MHD equilibrium  and the bootstrap 
current computations, one has to balance these two for convergency. This makes the VMEC run cannot achieve its highest accuracy  in the high bootstrap current percentage case, otherwise, the recalculated bootstrap current would differ from the original bootstrap current input. 

Note that $q_{\min}$ raises from below 2 to above 2 from Case 1 to 5 as shown in Fig. \ref{advq}.
We especially plot the total plasma and vacuum energies for $n=2$ modes for the five equilibria in Case 1-5 in 
Fig. \ref{adve}. 
One can see that, when $q_{\min}$ increases above 2, the $n=2$ mode stability improves substantially. 

We have also studied  the MHD ballooning stability conditions. 
The unstable regions in the poloidal flux range  for Case 3-5 are respectively as follows: $(0.36,0.59)$, $(0.38,0.68)$,
and $(0.61,0.75)$. One can see that the unstable region is shifted outward as the beta and bootstrap current fraction
 increase.
From Fig. \ref{advp} one can see that for Case 5 the unstable region lies actually at the foot of the internal transport barrier. 
This may be good for maintaining the peaked pressure profile. Also, since the magnetic shear is low, we expect the kinetic
effects can play a role. Therefore, we have not tried to further optimize the pressure profiles for ballooning mode stability.
Instead, it is proposed for future studies with kinetic and nonlinear turbulence effects included.

From these results for negative triangularity configuration, one can see that 
 the normalized beta  can reach as high as about twice the extended Troyon limit (Case 5) for low-$n$ modes. It also exceeds 
 the usual Troyon limit $4$(I/aB).
This high beta stability for negative triangularity configurations with high bootstrap fraction and high poloidal beta
is observed in various peaked pressure profiles, including with or without an internal transport barrier.
The reasons for  good stability features of the negative triangularity equilibria with  high bootstrap current fraction and
peaked pressure profile are as follows: First, from Figs. \ref{advp} and \ref{advq} one can see that the pressure
 gradient is mostly confined in the  negative magnetic shear region, which is self-generated by the high bootstrap current;  
 In the corresponding positive triangularity case, instead, 
 the rotation transform is significantly lower, leading to poor stability.
 Second, note that the negative
 triangularity configuration tends to have a bigger bad curvature volume. However, this happens mostly in the outer part
 of plasma column. This adverse feature is minimized by the peaked pressure profile, 
  which places the pressure gradient at a lower negative triangularity region.
  These contribute to the better
 stability features achieved for low-$n$ modes.
 
Note that the Troyon limit for the positive triangularity case was developed primarily based on the eigenvalue codes and the evaluation
of the Mercier/ballooning stability. This
method is extended to the negative triangularity cases, for example,  Ref. \cite{yu15}.  Reference  \cite{yu15} uses the KINX code.
Similar to the KINX code, we use the eigenvalue code: AEGIS. To use AEGIS is because it is an adaptive code, which is good for studying  marginal stability.   
Due to the importance of the current results we use the DCON code for double check. 
 AEGIS and DCON  are both radially adaptive codes, based on the  shooting from 
 the vicinity of magnetic axis to the plasma edge.
  Generally speaking, DCON can get the marginal stability, while the eigenvalue codes usually have the limitation on the lowest growthrate to be able to reach. Since AEGIS is an adaptive code, its detectable  lowest  growthrate is lower than the
nonadaptive codes. Therefore, the AEGIS and DCON results are reasonably close for stability calculations. 
We are especially interested in the high beta cases.  For Cases 4 and 5 we see the so-called internal ``zero crossing" phenomenon with DCON computation, which may indicate that the fixed-boundary modes are unstable, while AEGIS runs do not see the instabilities. But, our extra check shows that they do not converge with respect to how small the starting $\chi$ is  and how many sidebands are used in DCON runs themselves. 
The stability and instability appear alternatively (for example, with sidebands 10 is unstable, 12 is stable, and 14 is unstable again...). 
  The uncertainty on the internal mode ``zero crossing" in the high beta equilibria  with Mercier interchange mode being stable is not because the DCON is an unreasonable code, nor because the convergency has not been checked.
This indicates that even if the most stringent energy minimization code DCON is used for double check,  no reliable unstable kink modes are found.
 We believe that
this numerical uncertainty may be because they are too close to the marginal stability so 
the stability conditions become sensitive to the numerical parameters. Note that the ideal MHD theory  is inapplicable for   the
too small growthrate cases. The non-ideal MHD effects, like the finite Larmor radius effect, can play a significant role there. This remains for future studies.

 We have also studied the cases without internal transport barrier.  Figure \ref{advntpq1} shows the pressure and $q$ profiles for the case without internal transport barrier. The equilibrium parameters are as follows: 
 $B_T=2$T,  $I=1.2$MA,   $a=0.6m$, 
$\beta_N^{geo}=4.780$, $\beta_N^{Troyon}=5.016$,  $\av{\beta}=4.780$, $l_i = 0.41$, and $f_{bs} = 0.89$.  Its cross-section is given in Fig. \ref{advcross}b.
In this case $I/aB=1$ and $4 l_i (I/aB) = 1.64$. The normalized beta  $\beta_N^{geo}$ is about 3 times the extended Troyon limit.  We also find that the $n=1,2,3 $ external kink modes are stable in this case. The Mercier interchange modes 
also are largely stable, except in the vicinity of magnetic axis. But, the ballooning modes are unstable in the $\chi$ regions:
$(0.2841, 0.3120)$ and $(0.428,  0.983$). This also shows the good stability potential especially for low $n$
 external modes.

Since we obtained the results with the normalized beta exceeding the Troyon limit, we have been
 extra cautious to check if the Troyon limit is there for the positive triangularity case. This is a check to the code system, 
although we have observed the Troyon limit in extensive computing experiences using the AEGIS/DCON system in decades. We indeed verified the Troyon limit in the positive triangularity case. The details are as follows.
The negative triangularity equilibria are compared with the positive triangularity configuration with the same type of peaked pressure profiles. 
No interesting cases are found to exceed significantly the Troyon limit in the positive triangularity cases.
A typical equilibrium with positive triangularity $\delta = 0.4$ is described as follows: 
$B_T=2$T,  $I=1.2$MA,   $a=0.6m$, 
$\beta_N^{geo}=3.727$, $\beta_N^{Troyon}=3.310$,  $\av{\beta}=3.728$, $l_i = 0.62$, and $f_{bs} = 0.67$. Its cross-section is given in Fig. \ref{advcross}a
and pressure and q profiles are given in Fig. \ref{advptpq1}.
The stability results are given as follows:
The critical wall positions for $n=1,2$ external kink modes are  respectively $1.55$ and $1.45$; while the $n=3$ modes are no-wall stable.
Compared with the data in Table \ref{table1}, one can see that the positive triangularity equilibrium with  a peaked pressure profile
is more unstable than the negative one.   We have also studied the Mercier and ballooning stability for this positive triangularity case.
The system is largely Mercier stable, except in the vicinity of magnetic axis. But, the ballooning modes are unstable in the $\chi$ regions:
$(0.0792,0.0902)$ and $(0.2279,0.4937)$. 

The results show
 that the negative triangularity configuration  is favorable for steady-state confinement
 in the advanced tokamak scenario with high bootstrap current fraction, high poloidal beta, and peaked pressure profile. 
It can achieve an even higher normalized beta  than the positive triangularity 
case.  As shown in Fig. \ref{troyon}, in a certain parameter domain,
the normalized beta  can reach about twice as high as the so-called Troyon limit for positive triangularity tokamaks as given in Ref. \onlinecite{betand3d}. 
The beta value seems to be limited by the high $n$ ballooning modes for this type of configurations. 
Nevertheless, the ballooning instabilities occur only at the foot of the transport barrier or the peaked pressure profile. 
The results show that the advanced steady-state scenario of negative triangularity we developed is
 not only good for divertor design but also good for steady-state confinement and MHD stability.

%%%%%%
 \section{Conclusions and discussion}
 \label{conc}
 %%%%%%

In conclusion, in equilibrium, we found that  the negative triangularity tokamaks are more
effective in generating the magnetic field line rotation  transform,
which especially shows that
 the negative triangularity tokamak can be very attractive for a 
steady-state fusion reactor  \cite{ziaea,zint,zrot}. The advanced steady-state scenario reported in this work
can work about without an externally driven current while maintaining the safety factor near the plasma edge 
to be reasonably low. 

In stability, we report the results  both for the DIII-D-homologous configurations and the advanced steady-state scenario.
To compare with the advanced steady-state confinement scenario, we first review
the stability studies for the DIII-D-homologous equilibria \cite{ziaea,zint,zrot}.
It is  confirmed that the equilibrium in the 
DIII-D negative triangularity experiments with normalized beta  $2.6$ and L-mode profile
 is indeed stable to the low-$n$ MHD modes. We then discuss the three  triangularity cases
with the triangularity $\delta = 0.4$, $0.0$, and $-0.4$
 by extrapolating the DIII-D experimental configurations with L-mode profiles and  low bootstrap current fraction. 
 It is found that the positive triangularity case is the most stable,
the zero triangularity case remains in the middle, and the negative triangularity case is the most unstable 
 for low $n$ MHD modes. Nevertheless, the reduction of beta value 
for the negative triangularity configuration  is  not dramatic and is still within an acceptable range. 
Also, it is found that although the low $n$ modes are more unstable in the negative triangularity case,
 the intermediate-$n$ modes with a resistive wall are more stable in the negative triangularity case than the positive triangularity case \cite{zint}. This is consistent with the experimental observations 
 of the lower level of turbulent transport in the negative triangularity case in TCV and DIII-D experiments \cite{tcv,d3d}.
 We also pointed out  that
although the low $n$ modes are more unstable in the negative triangularity case,
the rotation and diamagnetic drift effects  are 
 more effective in the negative triangularity case than the positive triangularity case as shown in Ref. \onlinecite{zrot}. 

Most importantly, our work shows the promising prospects of negative triangularity tokamak  for advanced steady-state confinement of fusion plasmas with high beta.
We examine the negative triangularity configurations with high bootstrap
current fraction, large poloidal beta, and peaked pressure profiles. We find that
the negative triangularity configuration is more stable than the positive
triangularity case in this scenario. In a certain parameter domain, the normalized beta  limit
 can reach about $8$ $l_i (I/aB)$ for low-$n$ modes, about twice 
the extended Troyon limit $4l_i (I/aB)$ for the conventional positive triangularity tokamaks.   
The same type of computations confirms that the positive triangularity configuration
is indeed constrained by the Troyon limit. Only, the negative triangularity configuration
with  high bootstrap
current fraction, large poloidal beta, and peaked pressure profiles can substantially 
overcome the Troyon limit.
It is known in the earlier works, for example,  Ref. \cite{yu15}, 
that the positive triangularity case is good for H-mode with a broad pressure profile, 
but the negative triangularity case is not.
We find that the negative triangularity favors, instead, the peaked pressure profiles or with internal transport barrier. 
The good stability potential for negative triangularity configuration with high bootstrap
current fraction, large poloidal beta, and peaked pressure profiles results from that the pressure
 gradient is mostly confined in the  negative magnetic shear region, which is self-generated by the high bootstrap current. 

Our results indicate that the negative triangularity tokamaks are not only good for divertor design, but also can potentially achieve
high beta, steady-state confinement with a very high bootstrap current fraction. 
This steady-state scenario is good for tokamak fusion reactor. 
Since the high beta case is achieved with a peaked pressure profile or with an internal transport barrier,  
one can expect  that the performance is ELM-free. The low-$n$ mode stability also implies that the
 beta limit is determined by the high-$n$ modes and therefore the negative triangularity 
 discharge can have a reduced disruptivity.
Since the high beta case is related to the no-wall stability, one can expect a high resistive wall mode beta limit. Also,
the  peaked pressure profile, favored  in the negative
triangularity case, can be expected to give rise to  higher reactor fusion productivity.
 
 In the end, we would like to point out that
  the tokamak configuration optimization is an effort in decades and the current work is merely a step forward.
 More efforts  remain to be made in this field to further explore the full potential of the negative triangularity tokamaks. 
The current investigation is mainly based on the ideal MHD codes, although
we did show that the rotation and diamagnetic drift effects are more effective
in stabilizing the resistive wall modes in the negative triangularity case.
We have followed the conventional procedure in developing a new concept
by studying the low $n$ MHD modes and the Mercier/ballooning mode stability.
The results are double-checked with AEGIS and DCON codes.
 How the non-ideal MHD effects affect the
results remains for future studies. 

The authors would like to acknowledge Dr. Alan Turnbull for helpful discussions.
This research is supported by the Department of Energy Grants DE-FG02-04ER54742.

\newpage

Figure captions

\begin{figure}[htp]
\centering
\includegraphics[width=100mm]{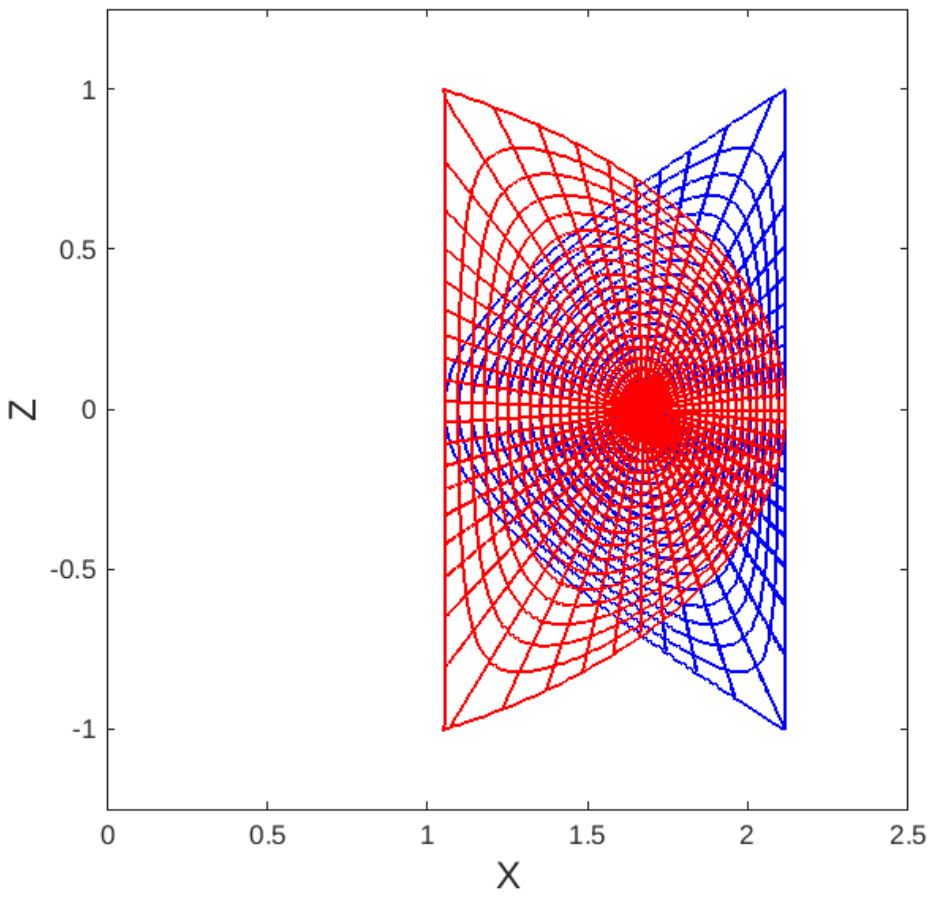}
\caption{~~}
\label{f2surf}
\end{figure}

Fig. \ref{f2surf}  The cross sections of positive  and negative triangularity configurations with the Solov\'ev solutions. The positive triangularity case is plotted
in red and negative triangularity in blue.

\begin{figure}[htp]
\centering
\includegraphics[width=90mm]{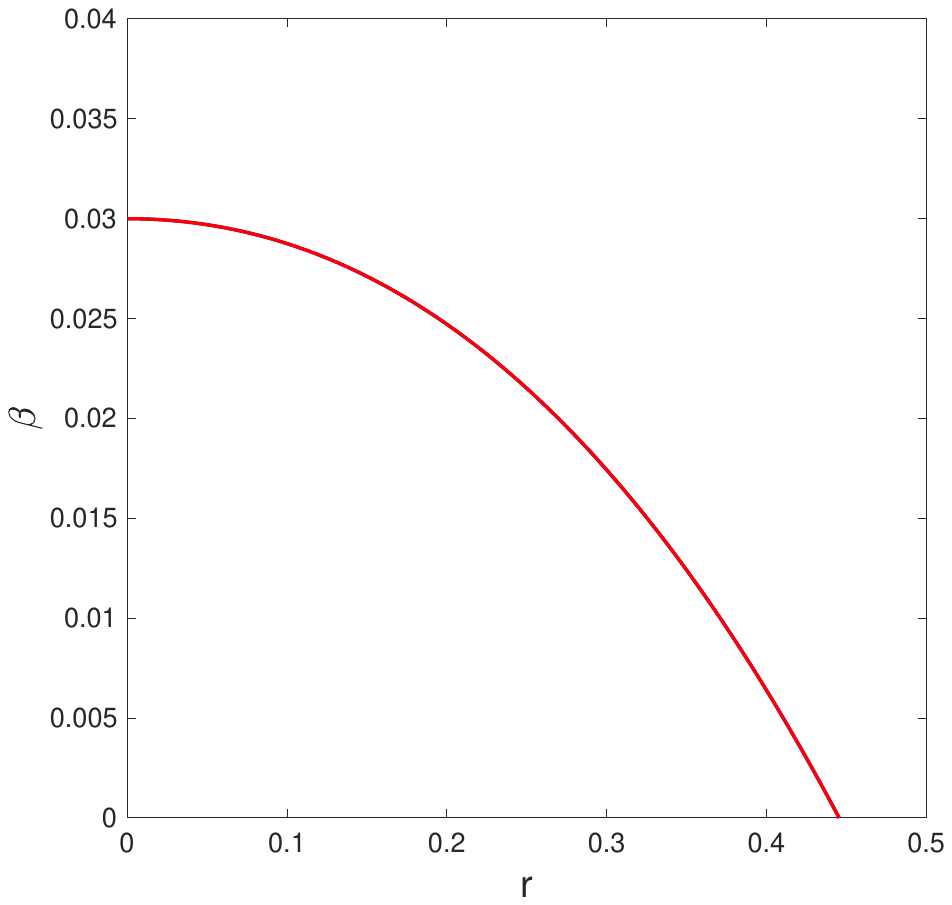}
\caption{~~}
\label{f2pr}
\end{figure}

Fig. \ref{f2pr}: The $\beta$ profiles versus the minor radius on the mid-plane for the positive  and negative triangularity configurations as described 
in Fig. \ref{f2surf}. The positive triangularity case is plotted
in red and negative triangularity in blue.

\begin{figure}[htp]
\centering
\includegraphics[width=90mm]{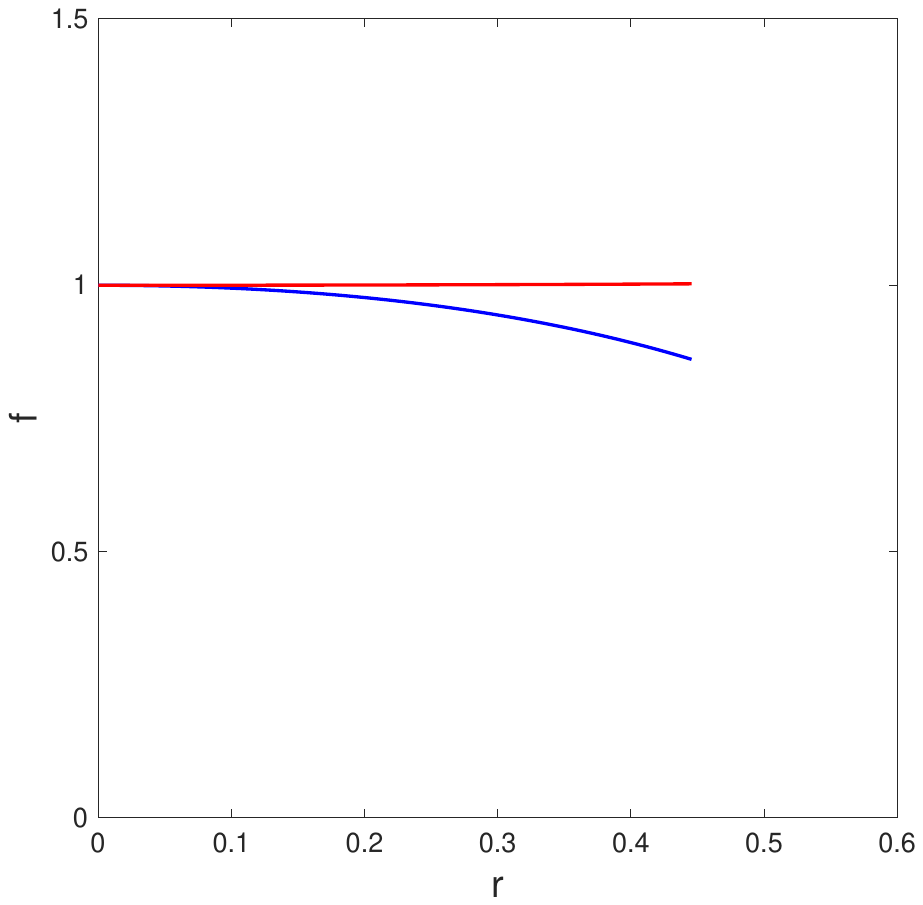}
\caption{~~}
\label{f2gt}
\end{figure}

Fig. \ref{f2gt}: The $f$ profiles versus the minor radius on the mid-plane for the positive  and negative triangularity configurations described 
in Fig. \ref{f2surf}. The positive triangularity case is plotted
in red and negative triangularity in blue.

\begin{figure}[htp]
\centering
\includegraphics[width=90mm]{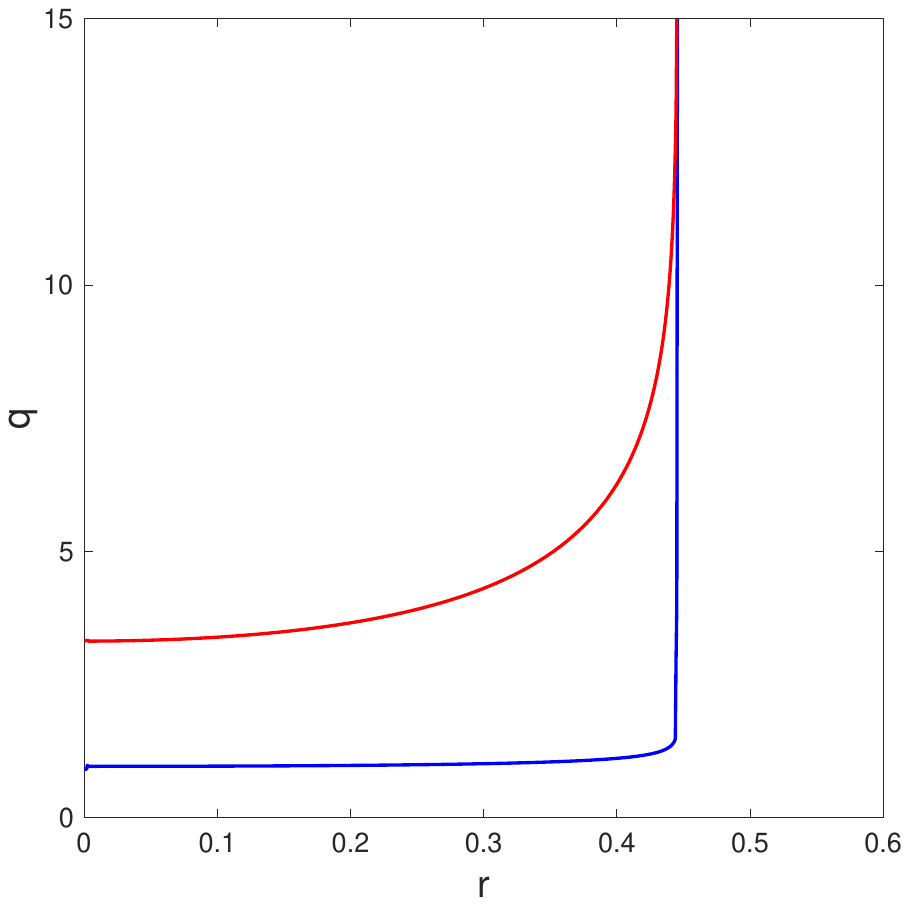}
\caption{~~}
\label{f2qsf}
\end{figure}

Fig. \ref{f2qsf}: The surface-averaged safety factor $q$ profiles versus the minor radius on the mid-plane for the positive  and negative triangularity 
configurations described 
in Fig. \ref{f2surf}. The positive triangularity case is plotted
in red and negative triangularity in blue.

\begin{figure}[htp]
\centering
\includegraphics[width=100mm]{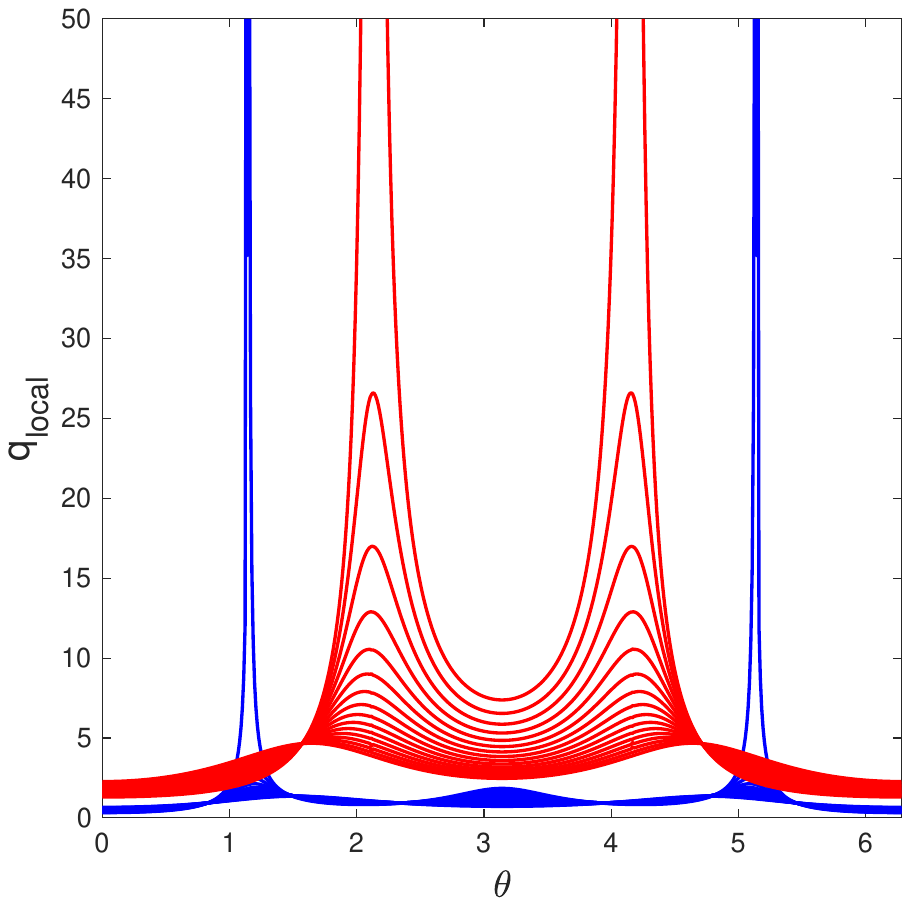}
\caption{~~}
\label{f2qloc}
\end{figure}

Fig. \ref{f2qloc}: The local safety factor ($q_{local}$) profiles versus the poloidal angle for the positive  and negative triangularity configurations as described 
in Fig. \ref{f2surf}. The positive triangularity case is plotted
in red and negative triangularity in blue. In each case, the consecutive $q_{local}$ are plotted from the plasma center to the edge.

\begin{figure}[htp]
\centering
\includegraphics[width=100mm]{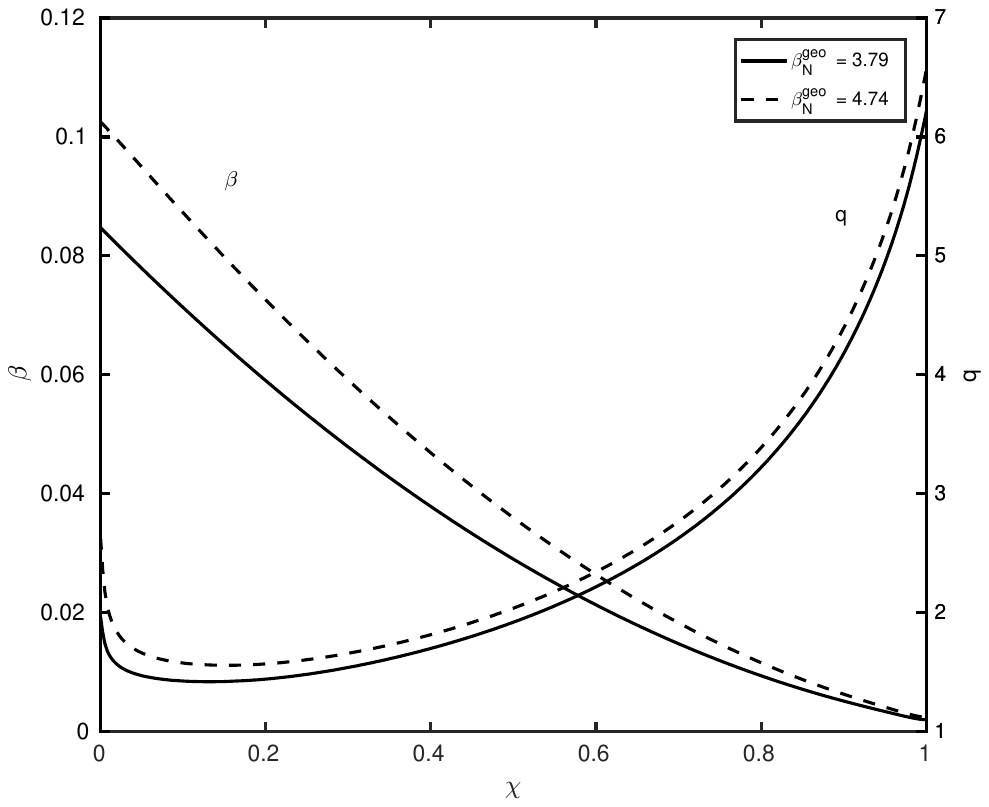}
\caption{~}
\label{dptpq}
\end{figure}

Fig. \ref{dptpq}. Plasma beta and safety factor profiles versus normalized poloidal magnetic flux $\chi$  for  DIII-D-homologous configuration with the triangularity $\delta =0.40$.

\begin{figure}[htp]
\centering
\includegraphics[width=100mm]{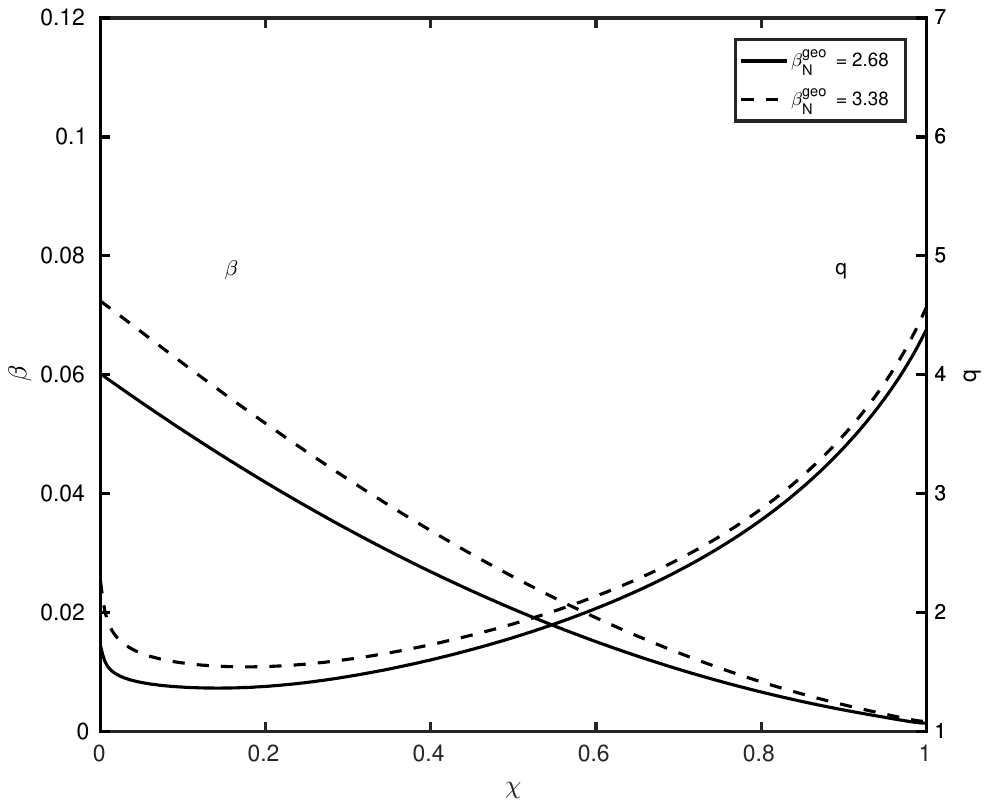}
\caption{~~}
\label{dntpq}
\end{figure}

Fig. \ref{dntpq}. Plasma beta and safety factor profiles versus poloidal magnetic flux for  DIII-D-homologous configuration with the triangularity  $\delta = - 0.40$.

\begin{figure}[htp]
\centering
\includegraphics[width=100mm]{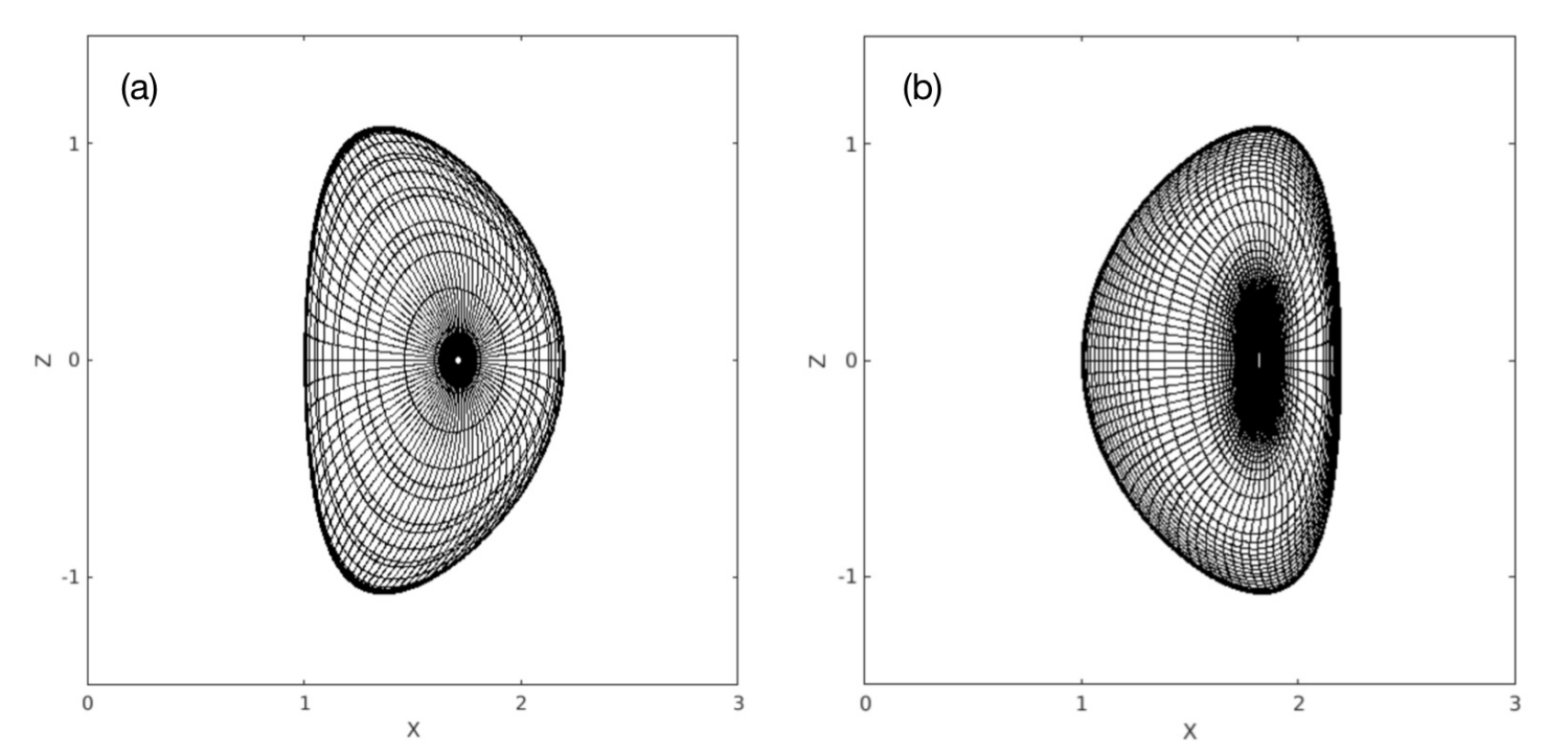}
\caption{~~}
\label{advcross}
\end{figure}

Fig. \ref{advcross}.  Cross sections of  positive and negative triangularity equilibria in the advanced steay-state
tokamak scenario with
high bootstrap current fraction.  The coordinate $X$ is
the major radius and $Z$ the height.  This figure has been presented in the 27th IAEA fusion energy conference  \cite{ziaea}.
The copy right for reusage is  based on the author right policy by the IOP publishing. 

\begin{figure}[htp]
\centering
\includegraphics[width=100mm]{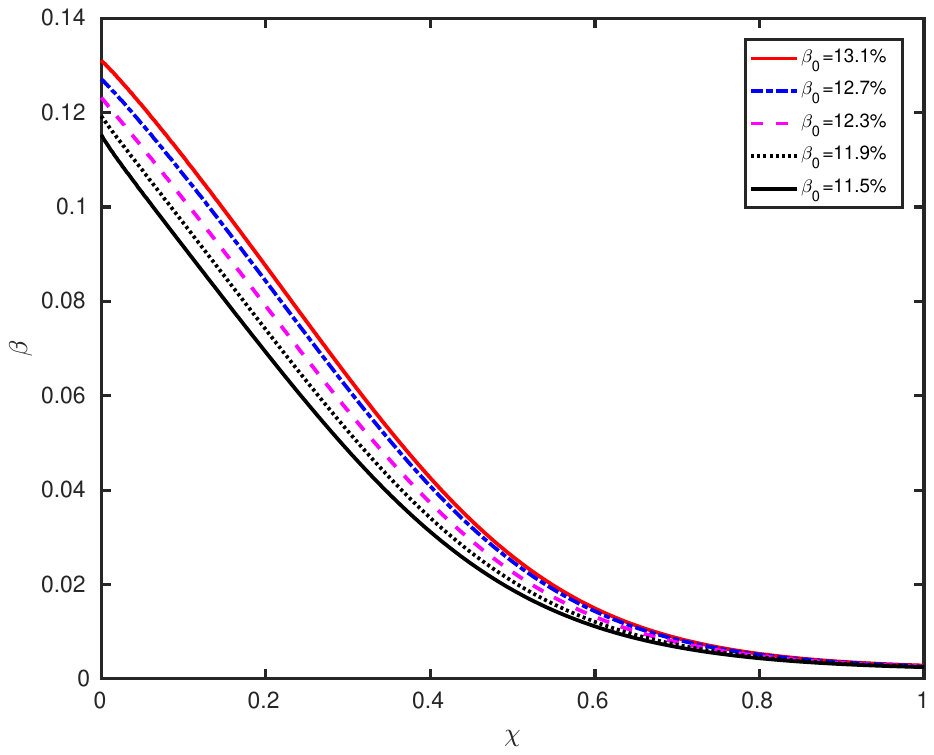}
\caption{~~}
\label{advp}
\end{figure}

Fig. \ref{advp}.  Pressure  profiles  versus poloidal magnetic flux for  negative triangularity equilibria
with the cross section shown 
in Fig. \ref{advcross}b in the advanced tokamak scenario with high bootstrap current fraction.

\begin{figure}[htp]
\centering
\includegraphics[width=100mm]{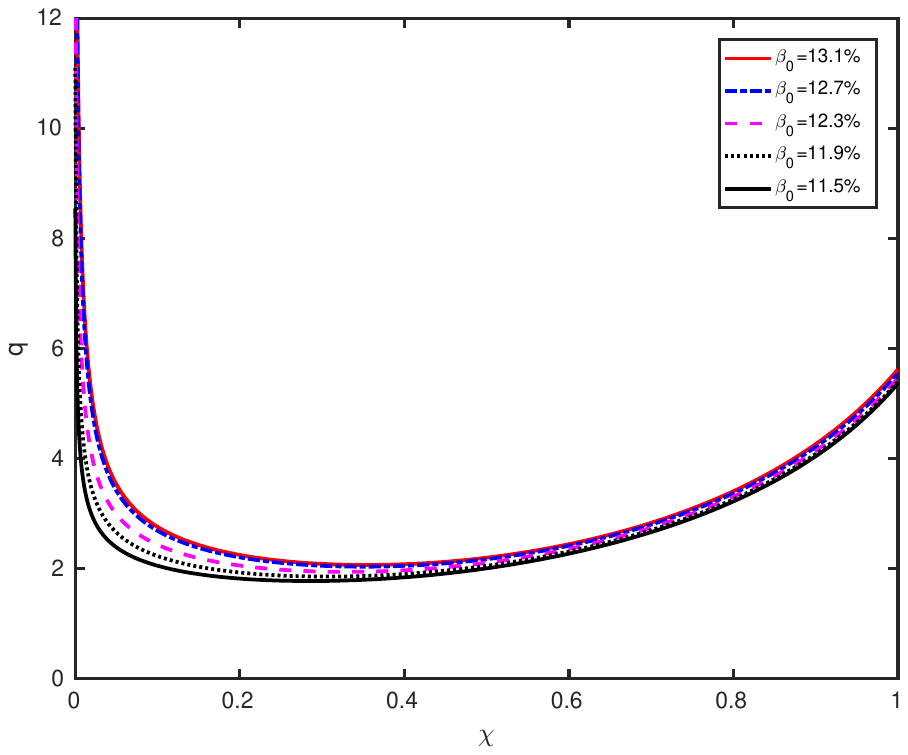}
\caption{~~}
\label{advq}
\end{figure}

Fig. \ref{advq}.  Safety factor profiles  versus poloidal magnetic flux for  negative triangularity equilibria with the cross section shown  
in Fig. \ref{advcross}b in the advanced tokamak scenario with high bootstrap current fraction.

\begin{figure}[htp]
\centering
\includegraphics[width=100mm]{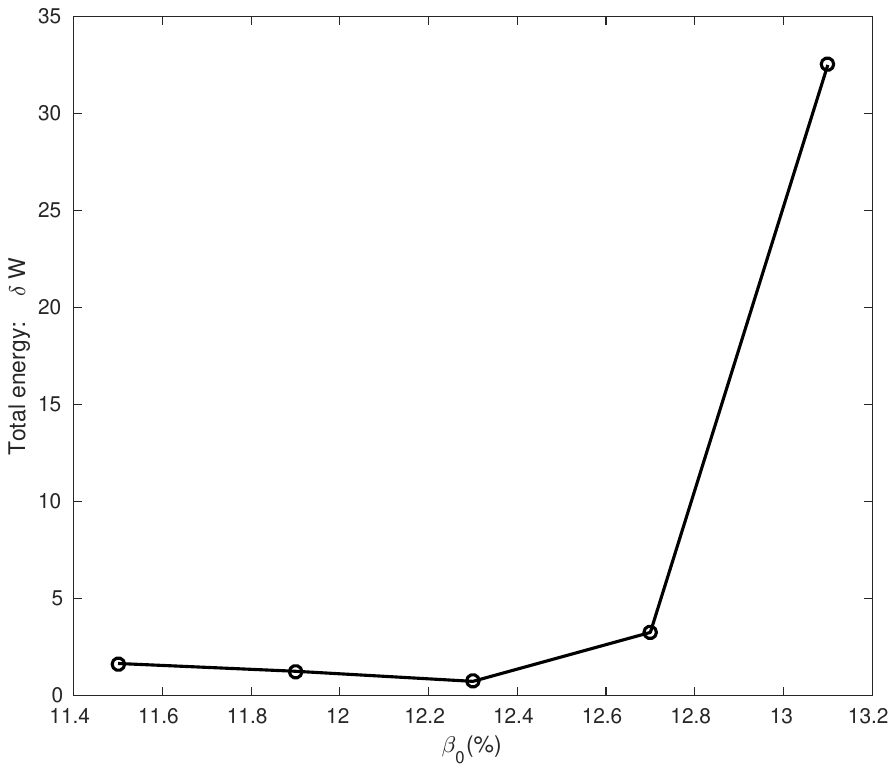}
\caption{~~}
\label{adve}
\end{figure}

Fig. \ref{adve}.  The total perturbation energies for $n=2$ modes for five equilibrium cases as given in Table \ref{table1} as computed by DCON.
A larger positive $\delta W$ corresponds to a better stability.

\begin{figure}[htp]
\centering
\includegraphics[width=100mm]{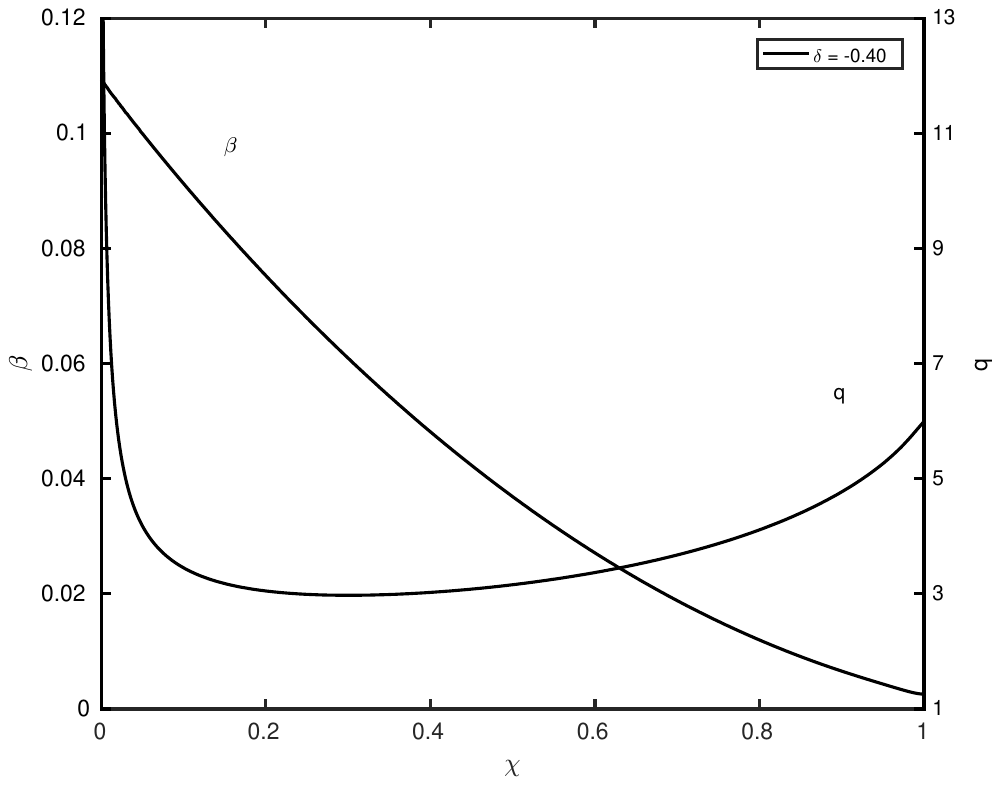}
\caption{~~}
\label{advntpq1}
\end{figure}

Fig. \ref{advntpq1}.  Plasma beta and safety factor profiles versus normalized poloidal magnetic flux with the 
triangularity  $\delta =  -0.40$ for the configuration with the cross section 
 shown in  Fig. \ref{advcross}b. The pressure profile is peaked, but does not have the internal transport barrier.

\begin{figure}[htp]
\centering
\includegraphics[width=100mm]{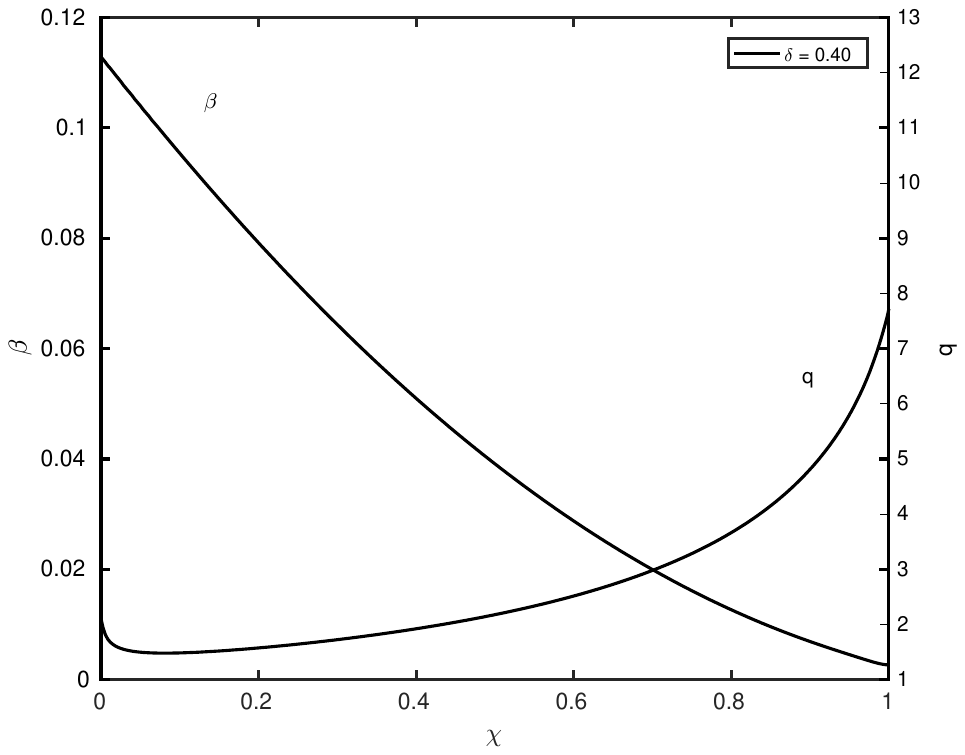}
\caption{~~}
\label{advptpq1}
\end{figure}

Fig. \ref{advptpq1}.  Plasma beta and safety factor profiles versus normalized poloidal magnetic flux with the
triangularity  $\delta =  0.40$ for the configuration with the cross section 
 shown in  Fig. \ref{advcross}a.

\begin{figure}[htp]
\centering
\includegraphics[width=80mm,angle=-90]{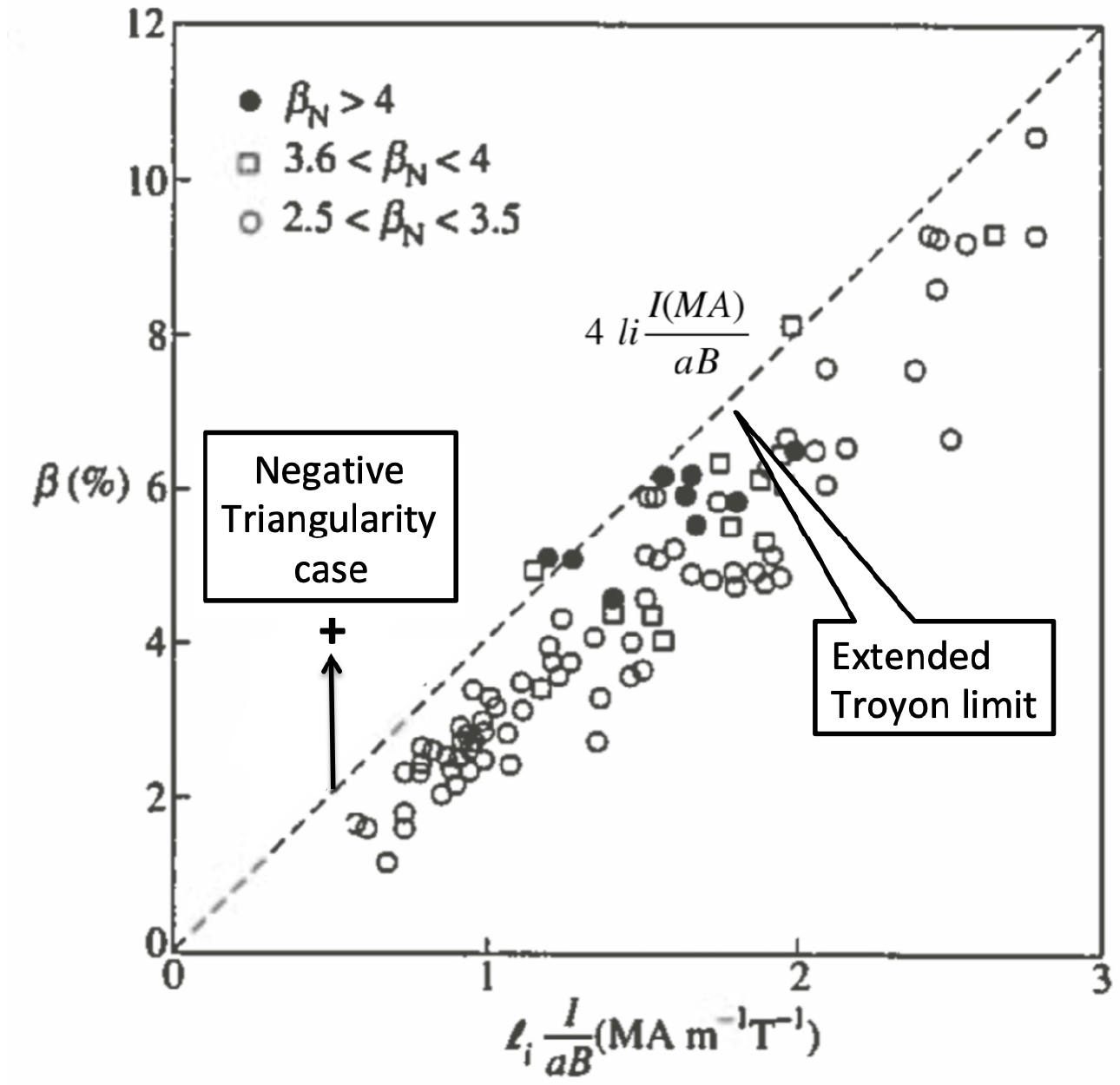}
\caption{~~}
\label{troyon}
\end{figure}

Fig. \ref{troyon}.  The normalized beta  limit for low $n$ external modes for negative triangularity tokamak configuration in the advanced steady-state scenario, as compared to the experimental observations for the positive triangularity cases (circles or squares). The dashed line is the so-called Troyon limit. Reproduced from Ref. \onlinecite{betand3d} with the negative triangularity results added.

\newpage

\newpage

\newpage

\newpage

\newpage

\newpage

\newpage

\newpage

\newpage

\newpage

\newpage

\newpage

\newpage

\newpage

\end{document}